\documentclass[a4paper]{article}
\usepackage{graphicx,amsmath,amssymb}
\newcommand{\haak}[1]{\!\left(#1\right)}
\newcommand{\rhaak}[1]{\!\left [#1\right]}

\newcommand{\gemr}[1]{\left. #1\right\rangle}

\newcommand{\lhaakl}[1]{\left |#1\right.}

\newcommand{\ket}[1]{\lhaakl{\gemr{#1}}}

\renewcommand{\imath}{\text{i}}

\newtheorem{assumption}{Assumption}

\begin{document}
\title{Changing the past by forgetting} 
\author{Saibal Mitra}
\date{\today}
\maketitle
\begin{abstract}
Assuming the validity of the Many Worlds Interpretation (MWI) of quantum mechanics, we show that memory erasure can cause one to end up in a different sector of the multiverse where the contents of the erased memory is different.

\end{abstract}
\section{Introduction}
As pointed out by David Deutsch \cite{deutsch}, it is possible to experimentally disprove all collapse interpretations of quantum mechanics if one could make measurements in a reversible way. Suppose an observer measures the $z$-component of a spin that is polarized in the $x$-direction. Then there exists a unitary operator that disentangles the observer from the spin, causing the observer to forget the result of the measurement. However, he would still remember having measured the $z$-component of the spin. In the MWI, the spin will be in its original state and therefore measuring the $x$-component will yield spin up with 100\% probability. In any collapse interpretation, measuring the $x$-component will yield spin up or spin down with 50\% probability.

Unfortunately, such reversible measurements are not possible with current technology. In the MWI, it is still the case that simply dumping the information about the result of the measurement in the environment will cause the observer to become disentangled from the spin, but now the spin is entangled with the rest of the universe. While subsequent measurements cannot distinguish the MWI from collapse interpretations, in the MWI the outcome of measuring the $z$-component again is not predetermined. In this article, we show how machine observers can benefit from resetting their memories to previous states. First, we state the assumptions that define what we mean by the MWI in this article.

\section{Assumptions}
The conclusions of this article depend only on the following assumptions about time evolution and the physical nature of the subjective experiences of observers. 
\begin{assumption}
The time evolution of the universe, including any observers, is always of form
\begin{equation}
\ket{\psi\haak{t_{2}}}=\hat{U}\haak{t_{2},t_{1}}\ket{\psi\haak{t_{1}}},
\end{equation}
where $\ket{\psi\haak{t}}$ represents the state of the universe at time $t$ and $\hat{U}$ is some unitary operator.
\end{assumption}

\begin{assumption}
The states any given observer can subjectively find herself in can be described classically using a finite amount of information, similar to specifying the computational state of a classical computer by specifying the state of each of the individual bits to be either zero or one. 
\end{assumption}
To describe the full quantum mechanical state vector of an observer will, of course, require a huge amount of additional information. In this article we will take the view that whatever the exact quantum mechanical state vector is, the possible states the observer's consciousness can be in, can be identified with some classically describable macrostates of the observer. We will consider the additional information needed to specify the exact quantum state of the observer as part of the rest of the universe. 

Denoting the macrostates an observer can find herself in formally as orthonormal ket vectors $\ket{O_{n}}$, the generic form of a quantum state of the universe containing the observer can be written as:
\begin{equation}\label{formm}
\ket{\psi} = \sum_{n,m}a_{n,m}\ket{O_{n}}\ket{\phi_{m}},
\end{equation}
where the $\ket{\phi_{m}}$ form a complete set of orthonormal states describing the rest of the universe. If we sum \eqref{formm} over $m$, we can write it formally as
\begin{equation}\label{form}
\ket{\psi} = \sum_{n}\ket{O_{n}}\ket{U_{n}}.
\end{equation}
The $\ket{U_{n}}$ can then be arbitrary states describing the rest of the universe. 

If an initial state
\begin{equation}
\ket{\psi_{\text{Initial}}}=\ket{O_{1}}\ket{U_{1}}
\end{equation}
evolves in time to become a superposition of the form
\begin{equation}
\ket{\psi_{\text{A while later}}} = \ket{O_{2}}\ket{U_{2}} + \ket{O_{3}}\ket{U_{3}},
\end{equation}
then we interpret this as two parallel universes, one containing the observer in the state $\ket{O_{2}}$, the other containing the observer in the state $\ket{O_{3}}$. 

\begin{assumption}
Born rule: In a state described by a superposition of the form
\begin{equation}
\sum_{k}\ket{O_{k}}\ket{U_{k}},
\end{equation}
the squared norms of the states $\ket{U_{k}}$ give the relative probabilities for the observer to find herself in the states $\ket{O_{k}}$.
\end{assumption}

\section{Memory erasure}
Consider a future machine observer who backs up its memory every day. It will reset its memory to the last backed up state when it learns about an impending disaster. The memory is also reset pseudo-randomly from macrostates that do not contain any information about a disaster. The fraction of such macrostates from which a resetting is done in the next clock cycle is $q$. Let's focus on the sector of the multiverse where at the time of a memory backup the observer is in some macrostate $\ket{O_j}$. The normalized state of the universe can be formally denoted as:
\begin{equation}\label{psi0}
\ket{\psi\haak{0}}=\ket{O_{j}}\ket{U_{j}}.
\end{equation}
This state then evolves to become at some clock cycle at time $t$, a state of the general form:
\begin{equation}\label{psit}
\ket{\psi\haak{t;j}}=\sum_{k}\ket{O_{k}}\ket{\tilde{U}_{k;j}}.
\end{equation}
Here the $j$ indicates that the information contained in the macrostate $\ket{O_{j}}$ has been stored. 
In the superposition described by \eqref{psit}, the observer in each of the macrostates $\ket{O_{k}}$ knows whether or not it will reset its memory in the next clock cycle. We split the summation in \eqref{psit} into three parts:
\begin{equation}
\ket{\psi\haak{t;j}}=\sum_{k_{1}}\ket{O_{k_{1}}}\ket{\tilde{U}_{k_{1};j}}+\sum_{k_{2}}\ket{O_{k_{2}}}\ket{\tilde{U}_{k_{2};j}} + \sum_{k_{3}}\ket{O_{k_{3}}}\ket{\tilde{U}_{k_{3};j}},
\end{equation}
where the summation over $k_{1}$ is over those values for which $\ket{O_{k_{1}}}$ will reset its memory because of a disaster, the summation over $k_{2}$ is over the states for which memory resetting is triggered by the pseudo-random generator, and the summation over $k_{3}$ is over states in which no memory resetting will happen at the next clock cycle.

Suppose that the probability for the observer to learn about an impending disaster during a clock cycle is $p$. The squared norm of the summation over $k_{1}$ is then $p$ and the squared norm of the sum of the summation over $k_{2}$ and $k_{3}$ is $\haak{1-p}$. Since a fraction $q$ of the macrostates that don't contain any information about a disaster will do a pseudo-random memory resetting, the squared norm of the second summation over $k_{2}$ is $\haak{1-p}q$. The probability for the observer to reset its memory is thus:
\begin{equation}\label{reset}
P_{\text{reset}} = p + \haak{1-p}q.
\end{equation}
The normalized wavefunction describing the sector of the multiverse containing the observer with its reset memory is given by:
\begin{equation}\label{resetst}
\ket{\psi\haak{t;j}}=\frac{1}{\sqrt{P_{\text{reset}}}}\ket{O_{j}}\rhaak{\sum_{k_{1}}\ket{\tilde{U}'_{k_{1};j}}+\sum_{k_{2}}\ket{\tilde{U}'_{k_{2};j}}},
\end{equation}
where the primes indicate that due to the dumping of the memory, the state of the rest of the universe has been modified. Since this process is unitary, it follows that upon making new observations, the observer with the reset memory will learn about a disaster coming its way with a probability of:
\begin{equation}\label{dis}
P_{\text{dis}} = \frac{p}{p + \haak{1-p}q}.
\end{equation}
The probability for the observer before memory resetting to ultimately learn of the disaster after the memory resetting is $P_{\text{reset}}P_{\text{dis}} = p$, so the memory resetting doesn't appear to have had any effect. Moreover, the above formulas for $P_{\text{reset}}$ and $P_{\text{dis}}$ are also valid in a purely classical setting. However, while one can give the probabilities a trivial classical interpretation, according to quantum physics, the outcome of the measurement by the observer of the state described by \eqref{resetst} is not pre-determined.

The only way to escape this conclusion is to assume that the stored state to which the memory resetting is done, is always different for the sectors in which a disaster has happened and in which it has not happened. This means that one has to assume that the state \eqref{psi0} can only evolve into a state where a disaster is certain to happen at the clock cycle at time $t$ or to a state in which it is certain that no disaster will happen at that time.

It is certainly the case that the different sectors of the multiverse where a disaster strikes and where it doesn't strike in the future will usually be already be very different some time before. Therefore, it is reasonable to expect that, due to decoherence, the information about the coming disaster has already affected the exact wavefunction of the observer. Nevertheless, this doesn't necessarily have to happen and even if it did, it is unreasonable to assume that the macrostate of the observer must then be affected, as that means that the observer's state of consciousness would necessarily have to differ in the two sectors.

\section{Discussion}
Assuming the validity of the MWI, we are forced to accept that by resetting the memory to a previous state, the reason why the memory was reset is no longer determined. In the limit $p\ll q$ the probabilities for memory resetting \eqref{reset} and for having to face a disaster after memory resetting \eqref{dis}, reduce to $P_{\text{reset}}=q$ and $P_{\text{dis}}=p/q$. The observer facing disaster can thus be almost sure to escape the disaster by doing a memory resetting. The observer who learns of a disaster after a memory resetting should think that he would not have faced the disaster had he not reset his memory.

Of course, none of this is directly verifiable to the observer as the observer only has access to information present in the sector of the multiverse where he resides. Although we cannot reset our memories, we often can choose to test or not to test for possible disasters. If the MWI is true then there is no benefit to test for impending disasters against which nothing can be done. After learning about the impending disaster all we can do is sit and wait for the disaster to happen, while we could be almost sure that we would not have to endure the coming disaster had we not tested for it in the first place. It was the detection of the impending disaster that trapped us in the wrong sector of the multiverse.

But perhaps when we forget something, this is equivalent to the memory resetting scenario discussed in this article. This depends on whether or not the lost memory has affected our consciousness. So, if we watch a recording of a soccer match played a long time ago, the outcome is undetermined, not just if we are watching the match for the first time and never read about the outcome, but perhaps also if we've seen the match before and forgot about the outcome.

\end{document}